\documentclass[reprint,amsmath,amssymb,aps,superscriptaddress]{revtex4-1}

\usepackage{graphicx}
\usepackage{hyperref}

\begin{document}

\title{Controllable Quantum Interference from Two-Photon Scattershot Sources}

\author{Joshua J. Guanzon}
\email{joshua.guanzon@uq.net.au}
\affiliation{Centre for Quantum Computation and Communication Technology, School of Mathematics and Physics, The University of Queensland, St Lucia, Queensland 4072, Australia}

\author{Austin P. Lund}
\affiliation{Centre for Quantum Computation and Communication Technology, School of Mathematics and Physics, The University of Queensland, St Lucia, Queensland 4072, Australia}

\author{Timothy C. Ralph}
\affiliation{Centre for Quantum Computation and Communication Technology, School of Mathematics and Physics, The University of Queensland, St Lucia, Queensland 4072, Australia}

\date{\today}

\begin{abstract}

We describe a multi-mode passive linear optical network which emulates the two-photon number statistics of a beam splitter, irrespective on where the two photons enter the network. This is done by firstly defining general properties that the generator of the network matrix must fulfil. We then show that the network's effective transmission coefficient can be freely set to replicate either the absence of coincidence counts (i.e. the Hong-Ou-Mandel dip), or a 100\% coincidence rate, as well as all possible two-photon beam splitter statistics between these two extremal points. Finally, we demonstrate that this network, in comparison to simpler systems, provides a better sampling rate and other resource advantages. 

\end{abstract}

\maketitle

\section{Introduction} \label{sec:MHD_intro}

In the past decade, the primary focus of quantum information processing has been on achieving exponential quantum advantages over classical computations. In this regard, linear optical networks coupled with single photon sources and detectors are a noteworthy platform, since they are experimentally accessible in comparison to other systems, and can in principle be used for universal quantum computation~\cite{knill2001scheme,yoran2003deterministic,nielsen2004optical,browne2005resource,ralph2005loss}. However, a key problem of this technology is the implementation of protocols which work at the very large scales required to approach universal computation. Recent experimental advances such as in integrated optical circuits, as well as improvements in theoretical protocols, have made implementing these large systems an increasingly practical reality~\cite{carolan2015universal,caspani2017integrated,qiang2018large,slussarenko2019photonic,bartlett2020universal}. A major theoretical advance came with the discovery of the boson sampling algorithm~\cite{aaronson2011computational}, which showed that classical computers can not efficiently simulate identical photons sent through a randomly chosen linear optical network~\footnote{The proof that is still used today is based on some assumptions about the distribution of matrix permanents for random matrices. However, these assumptions are highly plausible.}.

The aforementioned classical computation hardness argument was later expanded to encompass a problem called \emph{scattershot} boson sampling. This version of the problem considers the situation where the single photons sources are replaced with an array of non-linear crystals, which spontaneously generates entangled squeezed states that are used directly without any feed-forward mechanism~\cite{lund2014boson}. The scattershot source heralds multiple photons simultaneously with a heralding probability that reduces as the square-root of the number of photons. This source has the capacity to be hugely advantageous, and small scale experiments with this kind of device have already been successfully implemented~\cite{bentivegna2015experimental}. The trade-off is that the photons are distributed over many modes in a uniformly random fashion, though their exact location is made clear upon receiving the heralding signal. There is currently little known about the potential applications, beyond boson sampling, of this type of photon source.

\begin{figure}[htbp]
    \begin{center}
        \includegraphics[width=\linewidth]{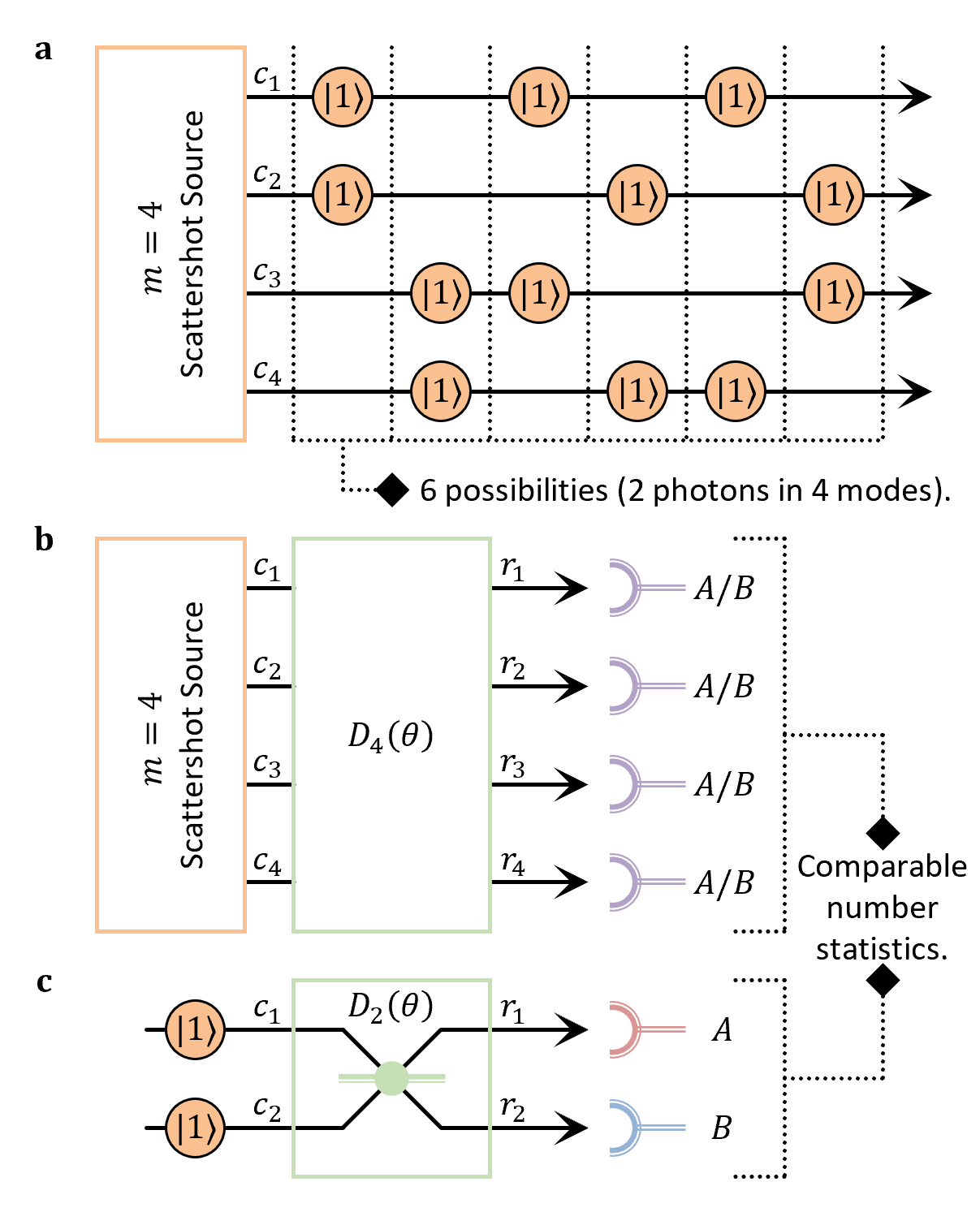}
        \caption{\label{fig:f1} 
            \textbf{a} The two-photon scattershot source injects two individual photons randomly into $m$ modes (here $m=4$). There are $m(m-1)/2$ possible input configurations. 
            \textbf{b} We investigate whether this random source can conduct some useful computation with a passive linear optical network $D_m(\theta)$ and $m$ photon detectors. 
            \textbf{c} We show there exists a $D_m(\theta)$ which passively replicates a beam splitter $D_2(\theta)$ two-photon interference effect for all possible input configurations and transmission ratios $\theta$, if we judiciously label the output detectors depending on the input.}
    \end{center}
\end{figure}

In this paper, we describe a passive optical network which is designed to take advantage of a two-photon variant of the scattershot source, and produces quantum interference effects in a predictable manner. This source  can be ideally modelled as an $m$ mode optical device which non-deterministically generates two separate, but otherwise indistinguishable, photons located anywhere amongst its $m$ channels $c_i^\dagger c_j^\dagger|0\rangle,$ $i\neq j\in[1,m]$. An example of the different possible configurations of the photons for $m=4$ is given in Fig.~\ref{fig:f1}\textbf{a}. Note that the interaction of two photons is the simplest non-trivial example of quantum interference, hence this study represents a necessary step towards understanding passive quantum processing with scattershot sources. 

We will show, for all possible inputs from an $m$ mode two-photon scattershot source, there exists an $m$ mode passive circuit $D_m(\theta)$ with a configurable parameter $\theta$, that can interfere two photons as if they were incident on a standard beam splitter with $\cos^2\theta$ intensity transmission and $\sin^2\theta$ intensity reflection. This quantum interference includes the extensively studied Hong-Ou-Mandel (HOM) dip effect~\cite{hong1987measurement}, which occurs when a beam splitter's transmission ratio is set to 50:50. This HOM effect is due to the destructive interference of two possible outcomes, resulting in the signature effect of zero probability that a coincidence count will occur (i.e. the two photons will never be detected separately). We will call all possible linear optical networks that can accept all inputs from two-photon scattershot sources and output the quantum interference of an arbitrary beam splitter as Multimode HOM Devices (MHDs), in which $D_m(\theta)$ is a particular subset of all these possible optical circuits. 

This problem has been partially solved only for the 50:50 beam splitter case (i.e. for one specific $\theta$ configuration of our network) using Sylvester interferometers, a type of Hadamard transformation~\cite{crespi2015suppression}. This was studied recently in the context of photon indistinguishably and suppression (or destructive interference) laws, and it has been shown that these interferometers produce HOM-like interference between two photons entering it from any pair of input modes~\cite{crespi2015suppression,dittel2016many,viggianiello2018experimental,viggianiello2018optimal}. In our work, we want to find a configurable network $D_m(\theta)$ that naturally generalises Sylvester interferometers, in the sense that it can reproduce the statistics of not only a 50:50 beam splitter with a balanced transmission coefficient, but any general beam splitter with a $\theta$ transmission coefficient. Therefore, $\theta$ can be thought of as a parameter which determines the amount of interference experienced by two separate photons, irrespective of where those photons enter our circuit $D_m(\theta)$. This research thus introduces a type of scalable multi-mode network that acts like a general beam splitter for particular Fock state inputs. General beam splitters are very useful since they are the vital building blocks for all types of optical circuits. As an additional benefit, the construction method presented here has the potential to leverage the efficient multi-mode sampling and photonic generation advantages provided by scattershot sources at large-scales.

We begin by firstly characterising the linear optical network $D_m(\theta)$ in Section~\ref{sec:MHD_char}, which includes defining the necessary properties of the associated generator and corresponding matrix representation. Next, in Section~\ref{sec:MHD_dip} we will use these properties to show the existence of two $\theta$ critical points, which correspond to 0\% and 100\% coincidence count probability between two groups of output detectors. In Section~\ref{sec:MHD_inter} we will then prove that, between these two critical points, the probability profile is a decreasing continuous function as the network's analogous transmission coefficient $\theta$ is increased, just like in a standard beam splitter. Finally, in Section~\ref{sec:MHD_comp} the resource advantages and costs of this model will be analysed in comparison to other more straightforward systems.

\section{Characterisation of the MHD} \label{sec:MHD_char}

\subsection{Passive Linear Optical Networks} 

A passive linear optical network with $m\in\mathbb{N}$ modes can be, in general, represented by an $m \times m$ unitary matrix $U_m$. This matrix maps the transformations of the bosonic annihilation operators of the input commencing modes to the output resulting modes $U_m\vec{c}=\vec{r}$. The network is passive and linear, hence the number of photons and energy are conserved; note that we will only consider the situation where no photon loss occurs. In this paper, we will focus on the matrix $U_m=D_m(\theta)$, whose components are dependent on a single real variable $\theta\in\mathbb{R}$. This variable will later be shown to be physically analogous to the transmission ratio of a standard beam splitter. 

The $m$ channel two-photon scattershot source can be attached to the commencing modes of the device $D_m(\theta)$, as shown for the $m=4$ case in Fig.~\ref{fig:f1}\textbf{b}. So that we can conduct statistical measurements of the network, we need to use $m$ number-resolving detectors on the resulting modes. After a measurement is performed, the output detectors are split into two equal $m/2$ sized groups (which corresponds to labelling them either $A$ or $B$). Note that the particular grouping can change depending on which input was heralded by the scattershot source; this point will be elucidated later on in this paper. This detector grouping allows us to compare the probability of, for example, two photons landing in the $A$ group of detectors in the $D_m(\theta)$ case with the typical beam splitter $D_2(\theta)$ case (which only has one $A$ detector), as shown in Fig.~\ref{fig:f1}\textbf{c}. 

\subsection{Properties of the Generator} 

We will show that the stated beam splitter like interference of a MHD follows as a natural consequence of just three conditions on the generators $Y_m$. These are $m \times m$ real matrices which generate possible MHDs through an exponential map 
\begin{align}
    D_m(\theta) = \exp(\theta Y_m). \label{eq:MHD_exp}
\end{align}
Crucially, $Y_m$ is not dependent on $\theta$ (or any other variables), therefore studying these generators is a straightforward avenue to gaining insight into $D_m(\theta)$. There are three requirements on these generators, which we will briefly motivate in the next paragraph:
\begin{enumerate}
    \item[A1.] Skew-symmetric matrix 
    \begin{equation}
        Y^T_m = -Y_m.
    \end{equation}
    \item[A2.] Same magnitude off-diagonal components
    \begin{equation}
        |y_{i,j}|=|y_{p,q}|, \hspace{5 mm} \forall i\neq j,p\neq q.
    \end{equation}
    \item[A3.] Orthogonal matrix 
    \begin{equation}
        Y_m Y_m^T = \mathbb{I}_m.
    \end{equation}
\end{enumerate} 
Note that $y_{i,j}$ is the component of $Y_m$ located at the $i$th row and $j$th column, while $\mathbb{I}_m$ is the $m \times m$ identity matrix.  

The A1 condition guarantees the associated $D_m$ transformations are orthogonal matrices; this property can arise naturally from physical considerations of energy and number conservation in lossless beam splitters~\cite{campos1989quantum}. On the other hand, the A2 and A3 conditions physically correspond to input invariant quantum interference. We will later show in Section~\ref{sec:MHD_dip_allm} that these two additional restrictions guarantees an analogous HOM dip for all possible number of modes or sizes of our network. 

Note that we could also consider complex matrices, with requirements A1 and A3 instead replaced with anti-Hermitian $Y^\dagger_m = -Y_m$, traceless $\mathrm{tr}(Y_m) = 0$, and unitary $Y_mY_m^\dagger = \mathbb{I}_m$ conditions, which gives rise to particular unitary transformations. However, we will only consider real matrices in this paper for the following reasons: this generalisation to complex matrices is not all encompassing, it will not result in all possible MHDs; for pedagogical reasons, the arguments which will be made later are easier to grasp with real matrices; and finally, the resultant $D_m$ real matrices can be converted to the many other possible MHD forms using similarity transformations. 

A familiar example representation of this special orthogonal algebra in the simplest $m=2$ case is given by 
\begin{equation}
    Y_2 = 
        \begin{pmatrix}
             0 & 1 \\
            -1 & 0 
        \end{pmatrix}, \label{eq:MHD_Y2} 
\end{equation}
which also happens to satisfy A2 and A3~\cite{zee2016group,schwichtenberg2015physics}. This can then be used to generate the following MHD 
\begin{equation}
    D_2(\theta) = \exp(\theta Y_2) = 
        \begin{pmatrix}
             \cos\theta & \sin\theta \\
            -\sin\theta & \cos\theta 
        \end{pmatrix},  
\end{equation}
whose form is the stereotypical rotation transformation, commonly associated with beam splitters if we take $\theta$ to be the transmission ratio~\cite{campos1989quantum,zee2016group}. In the next section, we will analyse the matrix structure for a generalised $m$ amount of modes.

\subsection{Structure of the Matrix Representation}

As a consequence of the three previously stated conditions, the generator $Y_m$ has the general matrix form 
\begin{align}
    Y_m &= \frac{1}{\sqrt{m-1}} 
        \begin{pmatrix}
            0 & s_{1,2} & \cdots & s_{1,m} \\
            -s_{1,2} & 0 & \cdots & s_{2,m} \\
            \vdots & \vdots & \ddots & \vdots \\
            -s_{1,m} & -s_{2,m} & \cdots & 0
        \end{pmatrix}, 
\end{align}
where $s_{i,j}=\pm1, \forall i\neq j$ are sign placeholders. Note that the orthogonality condition A3 imposes additional restrictions to the signs $s_{i,j}$ of the off-diagonal components; only those configurations in which the columns (and rows) of $Y_m$ are orthonormal to each other are allowed. 

We can use these conditions to also show the general matrix form of $D_m$ is  
\begin{align}
    D_m(\theta) &= \exp(\theta Y_m) = \cos \theta \mathbb{I}_m + \sin \theta Y_m, \\ 
    &= 
        \begin{pmatrix}
            \cos\theta & \frac{s_{1,2}\sin\theta}{\sqrt{m-1}} & \cdots & \frac{s_{1,m}\sin\theta}{\sqrt{m-1}} \\
            -\frac{s_{1,2}\sin\theta}{\sqrt{m-1}} & \cos \theta & \cdots & \frac{s_{2,m}\sin\theta}{\sqrt{m-1}} \\
            \vdots & \vdots & \ddots & \vdots \\
            -\frac{s_{1,m}\sin\theta}{\sqrt{m-1}} & -\frac{s_{2,m}\sin\theta}{\sqrt{m-1}} & \cdots & \cos \theta
        \end{pmatrix}. \label{eq:MHD_Dm}
\end{align}
Therefore, the form of $D_m$ consists of $\cos \theta$ in the diagonal entries and $\pm \sin \theta / \sqrt{1-m}$ in the off-diagonals, with the particular sign chosen in a manner which satisfies orthogonality. Despite already knowing alot about the matrix structure, it is not obvious how one goes about finding a particular $Y_m$ (and thus $D_m$) for an arbitrary $m$; we will next show a straightforward method of creating higher mode matrices from lower mode matrices.

\subsection{Constructing Large-Scale MHD Networks} 

It will be convenient for us to have an easy method of determining higher order MHD matrices due to the nature of the scattershot source. This is because, as will be properly quantified later in Section~\ref{sec:MHD_comp}, it is advantageous in terms of sampling rate to have an MHD network with as many modes as possible. If we know of a particular generator $Y_m$, we can calculate another matrix $Y_{2m}$ with double the amount of modes by inserting it into the following block matrix 
\begin{align}
    Y_{2m} = \frac{\sqrt{m-1}}{\sqrt{2m-1}}
        \begin{pmatrix}
            Y_m & Y_m + \frac{\mathbb{I}_m}{\sqrt{m-1}} \\
            Y_m - \frac{\mathbb{I}_m}{\sqrt{m-1}} & -Y_m 
        \end{pmatrix}. 
\end{align}
This construction approach is similar in spirit to the Sylvester construction method~\cite{crespi2015suppression}. However, through the exponential mapping given in Eq.~\eqref{eq:MHD_exp}, this method would ultimately result in a network that is configurable by the parameter $\theta$, in contrast to the fixed Sylvester interferometers.

We will now show that $Y_{2m}$ also satisfies the requirements of a generator. First, we note that it's transpose is equivalent to 
\begin{align}
    Y_{2m}^T &= \frac{\sqrt{m-1}}{\sqrt{2m-1}}
        \begin{pmatrix}
            Y_m^T & Y_m^T - \frac{\mathbb{I}_m }{\sqrt{m-1}} \\
            Y_m^T + \frac{\mathbb{I}_m}{\sqrt{m-1}} & -Y_m^T
        \end{pmatrix}, \nonumber \\ 
    &= \frac{\sqrt{m-1}}{\sqrt{2m-1}}
        \begin{pmatrix}
            -Y_m & -Y_m - \frac{\mathbb{I}_m}{\sqrt{m-1}} \\
            -Y_m + \frac{\mathbb{I}_m}{\sqrt{m-1}} & Y_m
        \end{pmatrix}, \nonumber \\
    &= -Y_{2m},
\end{align}
thus it also satisfies the skew-symmetric matrix A1 condition. By inspection, it is clear that $Y_m\pm\mathbb{I}_m/\sqrt{m-1}$ is a matrix where all of its components have a magnitude of $1/\sqrt{m-1}$. Hence the off-diagonals of $Y_{2m}$ all have the same magnitude of $1/\sqrt{2m-1}$, thus satisfying the A2 requirement where $|y_{i,j}|=|y_{p,q}|, \forall i\neq j,p\neq q$. Finally, we note that 
\begin{align}
    Y_{2m}Y_{2m}^T &= \frac{m-1}{2m-1}
        \begin{pmatrix}
            2\mathbb{I}_m + \frac{\mathbb{I}_m}{m-1} & 0_m \\
            0_m & 2\mathbb{I}_m + \frac{\mathbb{I}_m}{m-1}
        \end{pmatrix}, \nonumber \\
    &= \mathbb{I}_{2m},
\end{align}
hence we have shown that $Y_{2m}$ satisfies the orthogonality A3 condition. 

We have just shown that if we have a $Y_m$ that satisfies the generator conditions, we can easily double the amount of modes by calculating via the above block matrix method $Y_{2m}$, which is guaranteed by construction to also satisfy the generator conditions. For example, since we know of a $Y_2$ from Eq.~\eqref{eq:MHD_Y2}, we can readily calculate the associated $Y_4$ as follows
\begin{align}
    Y_4 = \frac{1}{\sqrt{3}}
        \begin{pmatrix}
             0 &  1 &  1 &  1 \\ 
            -1 &  0 & -1 &  1 \\
            -1 &  1 &  0 & -1 \\ 
            -1 & -1 &  1 &  0 
        \end{pmatrix}. 
\end{align}
Hence applying the exponential mapping gives 
\begin{align}
    D_4(\theta) = 
        \begin{pmatrix}
             \cos\theta & \frac{\sin\theta}{\sqrt{3}} & \frac{\sin\theta}{\sqrt{3}} & \frac{\sin\theta}{\sqrt{3}} \\ 
            -\frac{\sin\theta}{\sqrt{3}} & \cos\theta & -\frac{\sin\theta}{\sqrt{3}} & \frac{\sin\theta}{\sqrt{3}} \\
            -\frac{\sin\theta}{\sqrt{3}} & \frac{\sin\theta}{\sqrt{3}} & \cos\theta & -\frac{\sin\theta}{\sqrt{3}} \\ 
            -\frac{\sin\theta}{\sqrt{3}} & -\frac{\sin\theta}{\sqrt{3}} & \frac{\sin\theta}{\sqrt{3}} & \cos\theta 
        \end{pmatrix}, 
\end{align}
which matches the general form given in Eq.~\eqref{eq:MHD_Dm}. By iterating this calculation, we have a practical procedure of determining $D_m$ to as many channels as you would want, to powers of two $m=2^k, k\in\mathbb{N}$. It can be shown that it is not possible to make these matrices for dimensions that are odd or singly even; we suspect that it is possible for doubly even $m=4k$ number of modes, in which the powers of two are a subset. We note that $D_m(\theta)$ is equivalent to a Hadamard matrix at a particular $\theta$ value, therefore the suspicion that these networks exist only for orders of $m=4k$ is further supported by the Hadamard matrix conjecture~\cite{hedayat1978hadamard}. 

We are now ready to derive, from the defined properties, the physical consequences for a general $D_m$ circuit, with particular reference to the two photon interference experienced by a typical two level beam splitter $D_2$. The previously calculated $D_4$ case will also be used as a concrete example in the next few sections.

\section{Analysis of Critical Points} \label{sec:MHD_dip}

\subsection{Two Mode HOM Dip} 

\begin{figure}[htbp]
    \begin{center}
        \includegraphics[width=\linewidth]{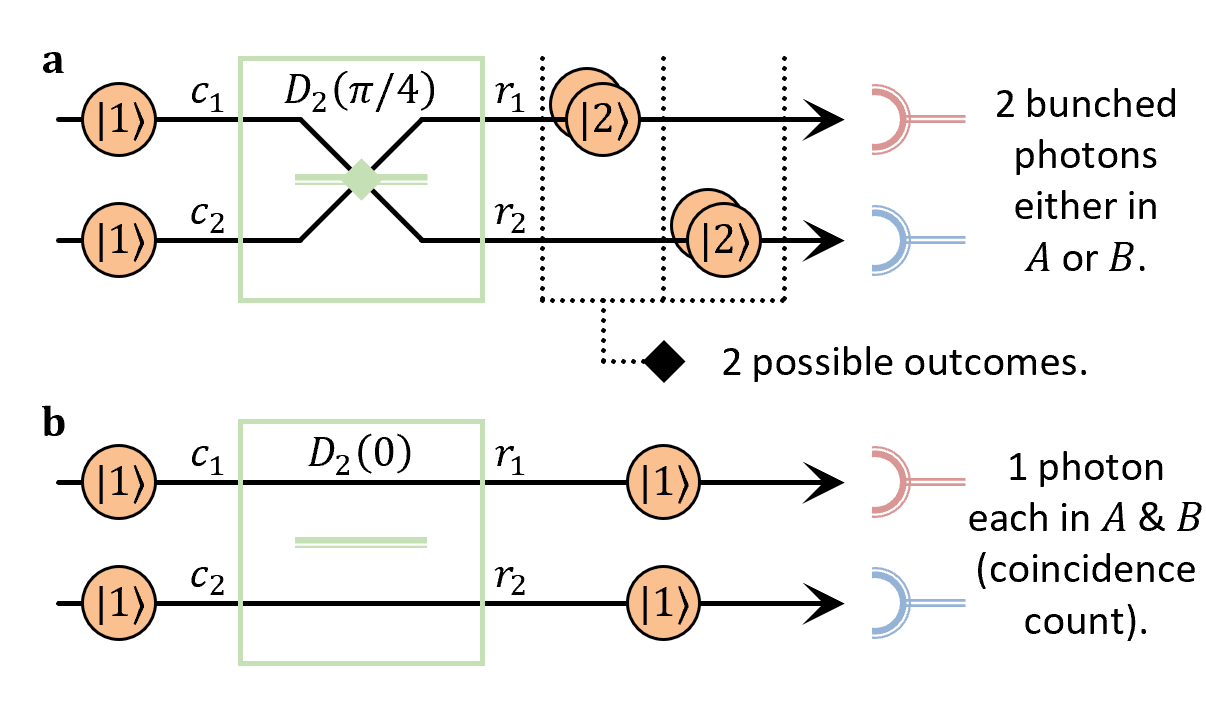}
        \caption{\label{fig:f2} 
            \textbf{a} A beam splitter, tuned to a balanced transmission $\theta_\mathrm{dip}(m=2)=\pi/4$, will have no coincidence counts. 
            \textbf{b} If a beam splitter is instead tuned to be essentially transparent $\theta=0$, it will always output coincidence counts.}
    \end{center}
\end{figure}

Suppose we have an input of two separate, but otherwise indistinguishable, photons into the two commencing modes $c_1^\dagger c_2^\dagger|0\rangle$ of a typical beam splitter $D_2(\theta)$. It can be shown that there exists a critical point $\theta_\mathrm{dip}(m=2)=\pi/4$ of the transmission ratio, where there is zero coincidence counts between the two output resulting modes $r_1^\dagger r_2^\dagger|0\rangle$. This HOM dip situation is summarised in Fig.~\ref{fig:f2}\textbf{a}. We can calculate the coincidence probability at $\theta_\mathrm{dip}$ formally as follows 
\begin{align}
    \cos(\theta_\mathrm{dip}) &= \sin(\theta_\mathrm{dip}) = \frac{1}{\sqrt{2}}, \nonumber \\ 
    D_2(\theta_\mathrm{dip}) &= \frac{1}{\sqrt{2}}
        \begin{pmatrix}
            1 & 1 \\
            -1 & 1 
        \end{pmatrix}, \nonumber \\ 
    \mathbb{P}_2(r_1 r_2 |c_1 c_2;\theta_\mathrm{dip}) &= \left|\mathrm{perm}\left[ \frac{1}{\sqrt{2}} 
        \begin{pmatrix}
            1 & 1 \\
            -1 & 1 
        \end{pmatrix} \right]\right|^2 = 0.
\end{align}
In this case, the probability amplitude is related to the permanent of the entire matrix~\cite{aaronson2011computational,scheel2004permanents}. Note that the permanent function is essentially the determinant but without the negative factors. 

\subsection{Coincidence Probability Between Two Groups of Detectors} 

The previous probability calculation can be extended generally to $D_m$. Suppose we want to know the probability of measuring a coincidence count between the \textit{resulting} modes $r_p^\dagger r_q^\dagger|0\rangle$, given an input into the system $D_m$ of two photons in the \textit{commencing} modes $c_i^\dagger c_j^\dagger|0\rangle$. Then the probability can be calculated by the squared permanent of a particular submatrix of $D_m$, composed of the intersection between \textit{rows} $\mathbf{r}_p$ and $\mathbf{r}_q$ with \textit{columns} $\mathbf{c}_i$ and $\mathbf{c}_j$ as follows
\begin{align}
D_{m}(\theta) &= \bordermatrix{ &   & \mathbf{c}_i &   & \mathbf{c}_j &   \cr
      & \cdot & \cdot & \cdot & \cdot & \cdot \cr
    \mathbf{r}_p & \cdot & d_{p,i} & \cdot & d_{p,j} & \cdot \cr
      & \cdot & \cdot & \cdot & \cdot & \cdot \cr  
    \mathbf{r}_q & \cdot & d_{q,i} & \cdot & d_{q,j} & \cdot \cr  
      & \cdot & \cdot & \cdot & \cdot & \cdot }, \\ 
\mathbb{P}_m(r_p r_q |c_i c_j;\theta) &= \left|\mathrm{perm}\left[ \begin{pmatrix}
d_{p,i} & d_{p,j} \\
d_{q,i} & d_{q,j}
\end{pmatrix} \right]\right|^2,
\end{align}
where we label the component in the $a$th row and $b$th column of $D_m$ as $d_{a,b}$~\cite{aaronson2011computational}. 

In modes higher than $m=2$, we need to introduce the notion of grouping the output detectors together for the analogous HOM dip to make sense. We can divide the $m$ resulting output modes into two equal $m/2$ sized sets $A(c_ic_j)=\{r_p,...,r_q\}$ and $B(c_ic_j)=\{r_u,...,r_v\}$, such that there is no overlap between them $A\cap B =\{\}$. Note that we allow the freedom where the particular detector grouping chosen depends on the input location of the two photons $c_i^\dagger c_j^\dagger|0\rangle$. We may then calculate the analogous coincidence probability between these two groups of detectors by 
\begin{equation}
    \mathbb{P}_m(A B |c_i c_j;\theta) = \sum_{r_a\in A} \sum_{r_b\in B} \mathbb{P}_m(r_a r_b |c_i c_j;\theta). 
\end{equation}
The analogous HOM dip condition occurs when the transmission coefficient is set to a particular $\theta=\theta_\mathrm{dip}$, which results in $\mathbb{P}_m(AB|c_ic_j;\theta_\mathrm{dip})=0$. Since probabilities can't be negative, this means we will need to show that each individual coincidence probability in the sum is zero 
\begin{equation}
   \mathbb{P}_m(r_a r_b |c_i c_j;\theta_\mathrm{dip}) = 0,\hspace{5 mm} r_a\in A, r_b \in B, 
\end{equation}
in which there are $(m/2)^2=m^2/4$ of these terms. Note that for the beam splitter $m=2$ case, it is clear with $A=\{r_1\}$ and $B=\{r_2\}$ that $\mathbb{P}_2(A B |c_1 c_2;\theta_\mathrm{dip}) =\mathbb{P}_2(r_1 r_2 |c_1 c_2;\theta_\mathrm{dip})=0$, as expected. 

\subsection{Higher Mode HOM Dip} \label{sec:MHD_dip_allm}

We will now show that an analogous HOM dip critical point exists at $\theta_\mathrm{dip}$, which we claim occurs when the transmission ratio is set to the following condition 
\begin{align}
\cos(\theta_\mathrm{dip}) &= \frac{\sin(\theta_\mathrm{dip})}{\sqrt{m-1}} = \frac{1}{\sqrt{m}}.
\end{align}
This is consistent with the previous $m=2$ case, in which this is the point where all the components have the same magnitude. Hence the corresponding matrix is given by  
\begin{align}
D_m(\theta_\mathrm{dip}) &= \frac{1}{\sqrt{m}} \begin{pmatrix}
1 & s_{1,2} & \cdots & s_{1,m} \\
-s_{1,2} & 1 & \cdots & s_{2,m} \\
\vdots & \vdots & \ddots & \vdots \\
-s_{1,m} & -s_{2,m} & \cdots & 1
\end{pmatrix}.
\end{align}
We note that this is essentially a skew-Hadamard matrix whose simplified structure is only possible because of the requirements on the generator. The components being the same magnitude, while potentially different signs, means that some of the contained $2\times2$ submatrices will have permanents that resolve to zero. In other words, the defined requirements on the generator, physically leads to the total destructive interference of probability amplitudes associated with certain coincidence outcomes. The precise details of this interference will now be explicitly elucidated. 

Let us suppose that the two input photons enter $D_m(\theta_\mathrm{dip})$ at two arbitrary commencing modes $c_i^\dagger c_j^\dagger|0\rangle$. The given output probabilities of these two photons are associated with only columns $\mathbf{c}_i$ and $\mathbf{c}_j$ of $D_m(\theta_\mathrm{dip})$ 
\begin{equation}
    \begin{pmatrix} 
    \mathbf{c}_i(\theta_\mathrm{dip}) & \mathbf{c}_j(\theta_\mathrm{dip})
    \end{pmatrix} = \frac{1}{\sqrt{m}} \begin{pmatrix} 
    s_{1,i}' & s_{1,j}' \\ 
    \vdots & \vdots \\
    s_{m,i}' & s_{m,j}' 
    \end{pmatrix},
\end{equation}
where $s_{u,v}'=\pm1, \forall u,v$. We note that there are $\binom{m}{2} = m(m-1)/2$ possible $2\times2$ submatrices contained within the $\begin{pmatrix} \mathbf{c}_i & \mathbf{c}_j\end{pmatrix}$ columns, whose permanents are proportional to the coincidence probabilities between two pairs of output modes.

We will now prove that an analogous HOM dip exists, by showing that $m^2/4$ of these $2\times2$ submatrices have permanents which resolve to zero (i.e. these correspond to the output modes pairs which have no coincidences). Let us look at the coincidence rate between two arbitrary output modes $r_p^\dagger r_q^\dagger|0\rangle$, which is related to the permanent $x$ as follows
\begin{align}
   x = \mathrm{perm}\left[ \frac{1}{\sqrt{m}} \begin{pmatrix}
s_{p,i}' & s_{p,j}' \\
s_{q,i}' & s_{q,j}'
\end{pmatrix}\right] &= \frac{1}{m}(s_{p,i}'s_{q,j}'+s_{q,i}'s_{p,j}') \nonumber \\ 
m s_{q,j}' s_{p,j}' x&= s_{p,i}'s_{p,j}' +s_{q,i}'s_{q,j}'. 
\end{align}
Note that we multiplied both sides by $m s_{q,j}' s_{p,j}'$, and then used the fact that $(s_{u,v}')^2=1,\ \forall u,v$. Notice that the permanent $x=0$, if $s_{p,i}'s_{p,j}' +s_{q,i}'s_{q,j}'=0$, or in other words $\mathrm{sign}(s_{p,i}'s_{p,j}') \neq \mathrm{sign}(s_{q,i}'s_{q,j}')$. We can now calculate how many $x$'s are zero by noting the fact that condition A1 means $D_m(\theta)$ are orthogonal matrices, hence each column must be orthogonal with each other for all values of $\theta$. Thus we know that 
\begin{equation}
    \mathbf{c}_i(\theta_\mathrm{dip}) \cdot \mathbf{c}_j(\theta_\mathrm{dip})= \frac{1}{m}(s'_{1,i} s'_{1,j} + \cdots + s'_{m,i} s'_{m,j}) = 0, 
\end{equation}
where we emphasise the fact that $s'_{p,i} s'_{p,j}=\pm1$. Since these terms add up to zero, it must be the case that $m/2$ of these terms are one $s'_{a,i} s'_{a,j}=1$, while the other $m/2$ of these terms are negative one $s'_{b,i} s'_{b,j}=-1$. This means that there are precisely 
\begin{equation}
    \left( \frac{m}{2} \right)^2 = \frac{m^2}{4}
\end{equation}
pairings in which $s_{a,i}'s_{a,j}' +s_{b,i}'s_{b,j}'=0$. Therefore $m^2/4$ of the $2\times2$ submatrix permanents are zero, and thus there is zero probability of detecting a coincidence count $\mathbb{P}_m(r_a r_b |c_i c_j;\theta_\mathrm{dip}) = 0$ between $m^2/4$ pairs of detectors, as needed to be shown. 

As an aside, we can now easily assign which output modes should correspond to which detector group (i.e. $A$ or $B$), using the simple calculation
\begin{equation}
    \mathrm{grp}(c_ic_j)=\mathrm{sign}(\mathbf{c}_i \odot \mathbf{c}_j),
\end{equation}
where $\odot$ is element-wise multiplication. Hence $\mathrm{grp}(c_ic_j)$ is a column vector with $m$ rows of $\pm1$ associated with each of the $m$ output detectors, which we then label $+1\rightarrow A$ and $-1\rightarrow B$. This condition is similar to proposed suppression laws in Sylvester interferometers~\cite{crespi2015suppression,dittel2016many,viggianiello2018experimental}. However, we will show in the next few sections that this same $A$/$B$ detector grouping could be used for all $\theta$ in our particular network, such that it leads to a similar probability profile as a general beam splitter.

\iffalse
\begin{figure}[htbp]
    \begin{center}
        \includegraphics[width=\linewidth]{p1.png}
        \caption{\label{fig:p1} 
            \textbf{a} The probability distribution of measuring particular output states, for all dispersed two photon inputs into $D_4(\theta_\mathrm{dip})$. 
            \textbf{b} With careful labelling of the detectors depending on the input (here for $c_2^\dagger c_3^\dagger |0\rangle=|0110\rangle$), one can always have no coincidence counts between the $A$ and $B$ detectors.}
    \end{center}
\end{figure}
\fi 

An example of $D_4$ at this critical point is given by
\begin{align}
    D_4(\theta_\mathrm{dip}) = \frac{1}{2}
        \begin{pmatrix}
             1 &  1 &  1 &  1 \\ 
            -1 &  1 & -1 &  1 \\
            -1 &  1 &  1 & -1 \\ 
            -1 & -1 &  1 &  1 
        \end{pmatrix}. 
\end{align}
Let us suppose the two-photon input is in the second and third modes $c_2^\dagger c_3^\dagger|0\rangle$. We should then label the output detectors according to the calculation
\begin{align}
    \mathrm{grp}(c_2c_3) = 
        \begin{pmatrix} 1 \\ 1 \\ 1 \\ -1 \end{pmatrix} \odot     
        \begin{pmatrix} 1 \\ -1 \\ 1 \\ 1 \end{pmatrix} =
        \begin{pmatrix} 1 \\ -1 \\ 1 \\ -1 \end{pmatrix}\rightarrow\begin{pmatrix} A \\ B \\ A \\ B \end{pmatrix}, 
\end{align}
which means resulting modes should be grouped as $A(c_2c_3)=\{r_1,r_3\}$ and $B(c_2c_3)=\{r_2,r_4\}$. We can then show that the coincidence probabilities between one $A$ labelled detector and one $B$ labelled detector is
\begin{align}
    \mathbb{P}_4(r_1 r_2 |c_2 c_3;\theta_\mathrm{dip}) &= \left|\mathrm{perm}\left[ 
    \frac{1}{2}\begin{pmatrix}
        1 & 1 \\
        1 & -1
    \end{pmatrix} \right]\right|^2 = 0, \nonumber \\
    \mathbb{P}_4(r_1 r_4 |c_2 c_3;\theta_\mathrm{dip}) &= \left|\mathrm{perm}\left[ 
    \frac{1}{2}\begin{pmatrix}
        1 & 1 \\
        -1 & 1
    \end{pmatrix} \right]\right|^2 = 0, \nonumber \\
    \mathbb{P}_4(r_2 r_3 |c_2 c_3;\theta_\mathrm{dip}) &= \left|\mathrm{perm}\left[ 
    \frac{1}{2}\begin{pmatrix}
        1 & -1 \\
        1 & 1
    \end{pmatrix} \right]\right|^2 = 0, \nonumber \\
    \mathbb{P}_4(r_3 r_4 |c_2 c_3;\theta_\mathrm{dip}) &= \left|\mathrm{perm}\left[ 
    \frac{1}{2}\begin{pmatrix}
        1 & 1 \\
        -1 & 1
    \end{pmatrix} \right]\right|^2 = 0. \nonumber 
\end{align}
Hence we have shown 
\begin{equation}
    \mathbb{P}_4(A B |c_2 c_3;\theta_\mathrm{dip}) = \sum_{r_a\in A} \sum_{r_b\in B} \mathbb{P}_4(r_a r_b |c_2 c_3;\theta_\mathrm{dip}) = 0, \nonumber 
\end{equation}
which means total destructive interference occurs between the two groups of detectors at $\theta_\mathrm{dip}$. This procedure can be repeated for all possible two photon inputs $c_i^\dagger c_j^\dagger |0\rangle$. In summary, we have shown that at the critical point $D_m(\theta_\mathrm{dip})$, we can split up the output mode detectors into two $m/2$ sized sets, such that there will never be a coincidence count between these two groups of detectors. 

\subsection{Two Mode 100\% Coincidence Rate} 

The two-level beam splitter has another critical point at $\theta=0$, where the two photons will always emerge in separate output modes $c_1^\dagger c_2^\dagger |0\rangle \rightarrow r_1^\dagger r_2^\dagger|0\rangle$. This can be calculated formally as follows 
\begin{align}
D_2(0) &= \begin{pmatrix}
1 & 0 \\
0 & 1 
\end{pmatrix}, \nonumber \\ 
\mathbb{P}_2(r_1 r_2 |c_1 c_2;0) &= \left|\mathrm{perm}\left[ \begin{pmatrix}
1 & 0 \\
0 & 1 
\end{pmatrix} \right]\right|^2 = 1.
\end{align}
This corresponds physically to setting the transmission coefficient to where the beam splitter is effectively transparent $D_2(0)=\mathbb{I}_2$, and thus does nothing to the input. Hence this critical point has a 100\% probability of measuring a coincidence count, as summarised in Fig.~\ref{fig:f2}\textbf{b}. 

\subsection{Higher Mode 100\% Coincidence Rate} 

We will show that, for all possible $D_m$, there exists a critical point where there is an analogous 100\% coincidence rate between the output detector groups $A$ and $B$. As with the $m=2$ case, setting the transmission coefficient of the device to $\theta=0$ results in the identity matrix 
\begin{align}
    D_m(0) &= \cos (0) \mathbb{I}_m + \sin (0) Y_m = \mathbb{I}_m.
\end{align}
At this critical point an input of two identical photons into two arbitrary modes $c^\dagger_i c^\dagger_j|0\rangle$ of $D_m(0)$ will always appear in the corresponding output modes $r^\dagger_i r^\dagger_j|0\rangle$. The probability of this outcome can be calculated as 
\begin{align}
    \mathbb{P}_m(r_i r_j |c_i c_j;0) &= \left|\mathrm{perm}\left[ \begin{pmatrix}
1 & 0 \\
0 & 1 
\end{pmatrix} \right]\right|^2 = 1,
\end{align}
where all other possibilities are zero. 

What we need to prove is that given a two photon input into $c_i$ and $c_j$, the corresponding modes $r_i$ and $r_j$ will always be in different detector groups. This means if $r_i \in A$, then we want to show that $r_j \in B$, so that we can say $\mathbb{P}_m(AB|c_i c_j;0)=\mathbb{P}_m(r_i r_j |c_i c_j;0)=1$. Now, to determine the grouping we look again at the HOM dip critical point $\theta=\theta_\mathrm{dip}$ where 
\begin{align}
    D_m(\theta_\mathrm{dip}) &= \frac{1}{\sqrt{m}}\bordermatrix{ &   & \mathbf{c}_i &   & \mathbf{c}_j &   \cr
      & \cdot & \cdot & \cdot & \cdot & \cdot \cr
    \mathbf{r}_i & \cdot & 1 & \cdot & s_{i,j} & \cdot \cr
      & \cdot & \cdot & \cdot & \cdot & \cdot \cr  
    \mathbf{r}_j & \cdot & -s_{i,j} & \cdot & 1 & \cdot \cr  
      & \cdot & \cdot & \cdot & \cdot & \cdot }. 
\end{align}
We know that $s_{j,i}=-s_{i,j}$ since A1 states that $Y_m$ is a skew-symmetric matrix. Now we note that  
\begin{align}
    \mathrm{sign}(\mathbf{c}_i \odot \mathbf{c}_j)_i &= s_{i,j}, \nonumber \\
    \mathrm{sign}(\mathbf{c}_i \odot \mathbf{c}_j)_j &= -s_{i,j}, \nonumber 
\end{align}
therefore by the grouping method we know that $r_i$ and $r_j$ must belong to different detector groups. Hence we have shown that for $D_m(0)$ we have a 100\% coincidence rate between the $A$ and $B$ detectors. Note that this result was determined using the skew-symmetric properties of our matrices, which is a property that doesn't hold for the matrices which describe Sylvester interferometers.

As a concrete example, consider again the $m=4$ case with $D_4(0) = \mathbb{I}_4$. For an input $c_2^\dagger c_3^\dagger |0\rangle$, we can calculate $\mathbb{P}_4(r_2 r_3 |c_2 c_3;0) = 1$, with all other probabilities being zero. We already previously determined that for an input of $c_2^\dagger c_3^\dagger |0\rangle$ we have $r_3\in A$ and $r_2\in B$. Therefore, we have shown that the total coincidence probability is $\mathbb{P}_4(AB|c_2 c_3;0)=\mathbb{P}_4(r_2 r_3 |c_2 c_3;0) =1$. This can be repeated for all possible inputs from a two-photon scattershot source. 

\section{Intermediate Transmission Values} \label{sec:MHD_inter}

We will now look at the number statistics between the two critical points of our network's configurable parameter $\theta\in[0,\theta_\mathrm{dip}]$. Recall, we want our network to reproduce the statistics of any arbitrary beam splitter with a $\cos^2\theta$ transmission. In other words, we will show that $\theta$ is an adjustable parameter which rationally controls the amount of interference for all possible pairs of input photons at the same time. The amount of interference will have a range between the two critical points covered in the previous section. This would mean this device would be useful beyond just the extremal values, as it could effectively be substituted in place of an arbitrary beam splitter in certain contexts, while being able to utilize the large-scale advantages of scattershot photonic sources.

\subsection{Two Mode Intermediate Transmission Values}

The coincidence probability can be calculated in the two mode case as follows 
\begin{align}
\mathbb{P}_2(r_1 r_2 |c_1 c_2;\theta) &= \left|\mathrm{perm}\left[ \begin{pmatrix}
\cos \theta & \sin \theta \\
-\sin \theta & \cos \theta 
\end{pmatrix} \right]\right|^2, \nonumber \\ 
&= \cos^2 (2\theta).
\end{align}
Note that between the two critical points $\theta\in[0,\pi/4]$, it is evident that $\mathbb{P}_2(r_1 r_2 |c_1 c_2;\theta)$ is a decreasing function from $\mathbb{P}_2(r_1 r_2 |c_1 c_2;0)=1$ to $\mathbb{P}_2(r_1 r_2 |c_1 c_2;\pi/4)=0$. This same decreasing property for the overall coincidence probability will be shown for all $D_m$. This property allows us to implement a one-to-one mapping between the number statistics of a typical beam splitter and the MHD. 

\begin{figure*}[htbp]
    \begin{center}
        \includegraphics[width=\linewidth]{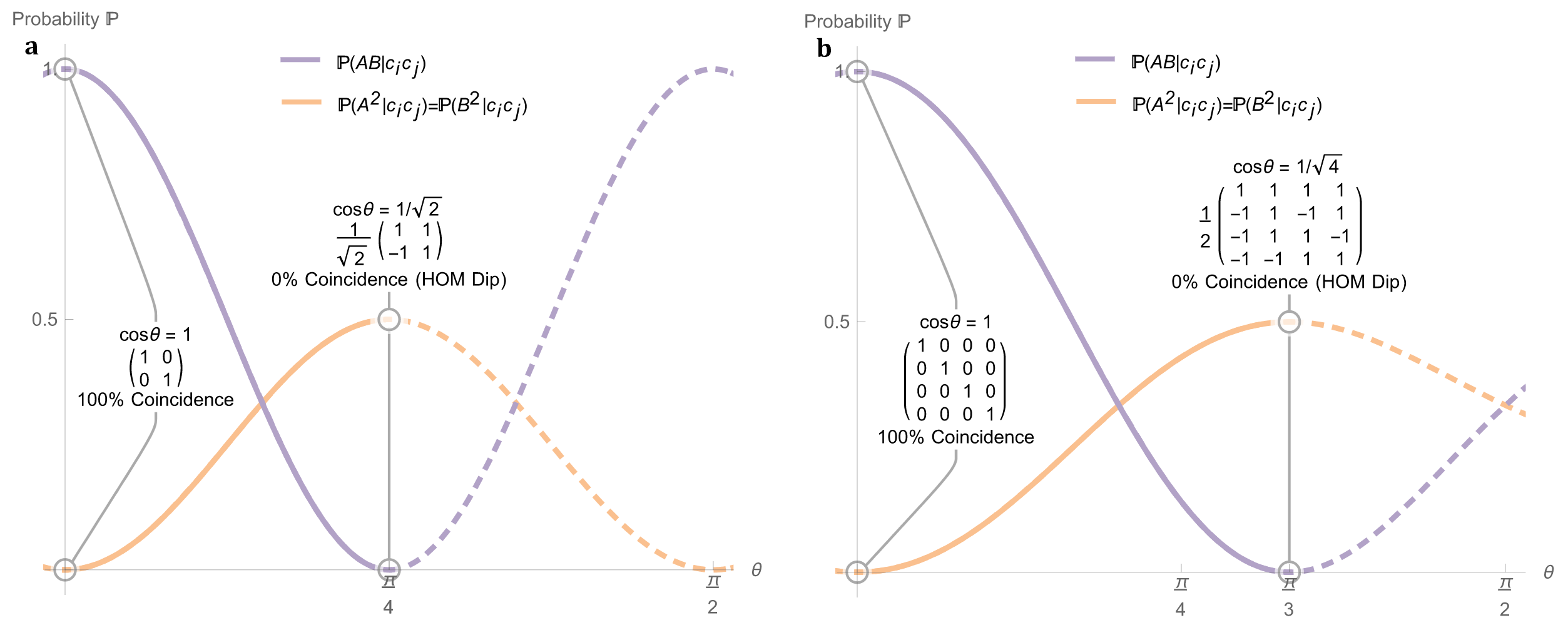}
        \caption{\label{fig:f3} 
            The number statistics associated with $A$ and $B$ labelled detectors, irrespective of where the two photons enter the system $c_i^\dagger c_j^\dagger|0\rangle$, for \textbf{a} the typical beam splitter $D_2(\theta)$, and \textbf{b} our particular linear optical network $D_4(\theta)$.} 
    \end{center}
\end{figure*}

\subsection{Higher Mode Intermediate Transmission Values} 

Suppose we have a two photon input into $D_m$, from separate arbitrary modes $c^\dagger_i c^\dagger_j|0\rangle $. The output probabilities only deals with columns the $\mathbf{c}_i$ and $\mathbf{c}_j$ of $D_m(\theta)$
\begin{equation}
    \begin{pmatrix} 
    \mathbf{c}_i(\theta) & \mathbf{c}_j(\theta)
    \end{pmatrix} = \begin{pmatrix} 
    \frac{s_{1,i}'}{\sqrt{m-1}}\sin \theta & \frac{s_{1,j}'}{\sqrt{m-1}}\sin \theta \\ 
    \vdots & \vdots \\
    \cos \theta & \frac{s_{i,j}'}{\sqrt{m-1}}\sin \theta \\
    \vdots & \vdots \\
    \frac{s_{j,i}'}{\sqrt{m-1}}\sin \theta & \cos \theta \\
    \vdots & \vdots \\
    \frac{s_{m,i}'}{\sqrt{m-1}}\sin \theta & \frac{s_{m,j}'}{\sqrt{m-1}}\sin \theta 
    \end{pmatrix}. \label{eq:MHD_cicj} 
\end{equation}
We will now examine the various forms of individual coincidence probabilities $ \mathbb{P}_m(r_a r_b |c_i c_j;\theta)$ where $r_a\in A$ and $r_b \in B$, which contribute to the total coincidence probability $\mathbb{P}_m(A B |c_i c_j;\theta)$ between detector groups $A$ and $B$. Note that the individual coincidence probabilities relate to the squared permanent of two rows in the previous matrix. There are $(m/2-1)^2$ coincidence probabilities with the particular form 
\begin{align}
    \mathbb{P}_m(r_a r_b |c_i c_j;\theta) &= \left( s_{a,i}'s_{b,j}'\frac{\sin^2 \theta}{m-1} + s_{b,i}'s_{a,j}'\frac{\sin^2 \theta}{m-1} \right)^2, \nonumber \\  
    &= 0,\hspace{5 mm} \forall a\neq i,b\neq j, \nonumber 
\end{align}
where we note that since these two rows belong to different output detector groups, it must be the case that $ s_{a,i}'s_{b,j}'=- s_{b,i}'s_{a,j}'$. There are also $m/2-1$ coincidence probabilities of the form
\begin{align}
    \mathbb{P}_m(r_i r_b |c_i c_j;\theta) &= \left(s_{b,j}'\frac{\cos \theta \sin \theta}{\sqrt{m-1}} + s_{i,j}'s_{b,i}'\frac{\sin^2 \theta}{m-1} \right)^2, \nonumber \\ 
    &= \left(\frac{\cos \theta \sin \theta}{\sqrt{m-1}} - \frac{\sin^2 \theta}{m-1} \right)^2,\hspace{5 mm} \forall b\neq j. \nonumber 
\end{align}
Note that $\mathbb{P}_m(r_a r_j |c_i c_j;\theta)$, $\forall a\neq i$, has the same form as the previous expression, where there are also $m/2-1$ of these coincidence probabilities. Finally, there is only $1$ coincidence probability of the form
\begin{align}
    \mathbb{P}_m(r_i r_j |c_i c_j;\theta) &= \left(\cos^2 \theta +s_{i,j}'s_{j,i}'\frac{\sin^2\theta}{m-1}\right)^2, \nonumber \\ 
    &= \left(\cos^2 \theta - \frac{\sin^2\theta}{m-1}\right)^2.\nonumber 
\end{align}
In total, these three forms represent the $(m/2-1)^2+2(m/2-1)+1=m^2/4$ individual coincidence probabilities between the $m/2$ detectors labelled $A$ and the $m/2$ detectors labelled $B$. Hence the total coincidence probability between the two detector groups is given by the following expression 
\begin{align}
     \mathbb{P}_m(A B |c_i c_j;\theta) &= (m-2)\left(\frac{\cos \theta \sin \theta}{\sqrt{m-1}} - \frac{\sin^2 \theta}{m-1} \right)^2 \nonumber \\ 
     &\hspace{5 mm}+ \left(\cos^2 \theta - \frac{\sin^2\theta}{m-1}\right)^2. \label{eq:MHD_pAB}
\end{align}
From this expression it is clear that setting the transmission ratio to $\theta=0$ results in a total coincidence rate of $\mathbb{P}_m(A B |c_i c_j;0) = 1$, in contrast the restriction $\cos(\theta_\mathrm{dip})=\sin(\theta_\mathrm{dip})/\sqrt{m-1}$ results in $\mathbb{P}_m(A B |c_i c_j;\theta_\mathrm{dip}) = 0$. Both these results are consistent with our previous analysis of the two critical points. We can now take the derivative of this total probability and show for all values of $m$ that 
\begin{align}
    \frac{d}{d\theta}\mathbb{P}_m(A B |c_i c_j;\theta) < 0,\hspace{5 mm} \theta\in(0,\theta_\mathrm{dip}).  
\end{align}
Hence we have shown that the total coincidence probability decreases, with increasing transmission coefficient, between these two critical points. 

The coincidence probability profile for $m>2$ is evidently not a facsimile of the $m=2$ typical beam splitter profile, as made clear in Fig.~\ref{fig:f3}. However, we have the freedom to move the parameter $\theta$ of $D_m$ however we like; since we know that these profiles always start at $1$, then decreases smoothly to $0$, we can easily set up a one-to-one mapping between any two probability profiles. Explicitly, we can set the following
\begin{align}
    \mathbb{P}_2(A B |c_1 c_2;\phi) &= \mathbb{P}_m(A B |c_i c_j;\theta), \\ 
    \cos^2 (2\phi) &= (m-2)\left(\frac{\cos \theta \sin \theta}{\sqrt{m-1}} - \frac{\sin^2 \theta}{m-1} \right)^2 \nonumber \\ 
    &\hspace{5 mm}+ \left(\cos^2 \theta - \frac{\sin^2\theta}{m-1}\right)^2,
\end{align}
such that all we have to do is numerically solve for $\phi(\theta)$ or $\theta(\phi)$. These functions will tell us how to move our network's parameter with respect to a typical beam splitter's transmission coefficient, such that their associated probability profiles will match exactly. Finally, we note that for high $m$ values, we can approximate the probability profile as 
\begin{align}
    \lim_{m\rightarrow\infty} \mathbb{P}_m(A B |c_i c_j;\theta) &= \cos^2\theta\sin^2\theta + \cos^4 \theta, \nonumber \\ 
    &= \cos^2 \theta. 
\end{align}
This means, to a good approximation for high amount of modes, we can fairly easily reproduce the beam splitter profile by moving the MHD parameter according to the simple expression $\theta(\phi)=2\phi$. 

\subsection{Two Photon Bunching Probability}

We note that there is symmetry associated with the two column matrix in Eq.~\eqref{eq:MHD_cicj}, where the rows are equally partitioned into the $A$ or $B$ groups (recall that the two unique rows containing the $\cos\theta$ diagonal elements are always in separate groups). This means the probability associated with detecting two photons in the $A$ modes will be the same as detecting two photons in the $B$ modes,
\begin{align}
    \mathbb{P}_m(A^2|c_i c_j;\theta) &= \mathbb{P}_m(B^2|c_i c_j;\theta). 
\end{align}
It must be the case that all possible probabilities add up to one, hence 
\begin{align}
    \mathbb{P}_m(A^2|c_i c_j;\theta)+\mathbb{P}_m(AB|&c_i c_j;\theta)+\mathbb{P}_m(B^2|c_i c_j;\theta) = 1, \nonumber \\
    \mathbb{P}_m(A^2|c_i c_j;\theta) &= \frac{1-\mathbb{P}_m(AB|c_i c_j;\theta)}{2}. \label{eq:MHD_pA2}
\end{align}
This was independently verified in the appendix by considering, once again, all the relevant permanents whose squared values will contribute to $\mathbb{P}_m(A^2|c_i c_j;\theta)$. 

\section{Resource Analysis} \label{sec:MHD_comp}

\begin{figure*}[htbp]
    \begin{center}
        \includegraphics[width=\linewidth]{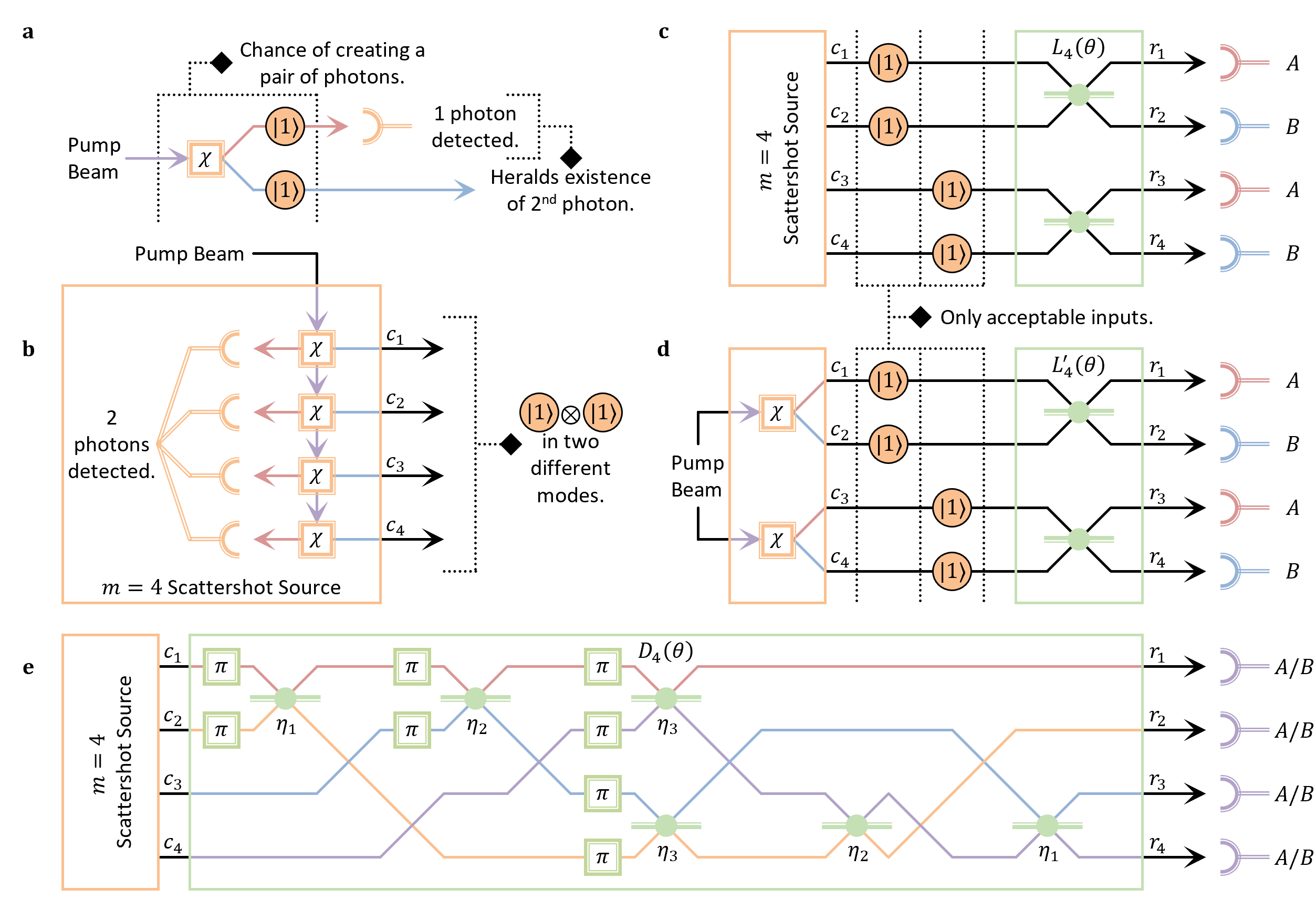}
        \caption{\label{fig:f4} 
            \textbf{a} A single $\chi$ squeezer crystal has a $(1-\chi^2)\chi^2$ chance of generating paired photons $|1\rangle\otimes|1\rangle$. If we detect 1 photon in the upper mode, this means another photon exists in the lower mode. 
            \textbf{b} By aligning an $n$ array of these $\chi$ crystals (here $n=4$), and accepting only the detection of $2$ individual photons in the left-most modes, means the output will effectively be $2$ separate photons located in the corresponding right-most modes. 
            \textbf{c} We compare our $D_m$ to a simple linear array of beam splitters $L_m$, which can only accept $n/2$ possible inputs. 
            \textbf{d} Another possible configuration for the source is attaching both arms of each crystal directly to the input of $L_m'$. This has the advantage of needing just $n=m/2$ crystals, however it only has $n$ acceptable inputs.
            \textbf{e} Circuit decomposition of $D_4$, composed of beam splitters with $\eta_i$ transmission ratio and $\pi$ phase shifters.} 
    \end{center}
\end{figure*}

Outside of theoretical interest, this device does provide in principle experimental advantages over more straight forward systems, such as a simple array of beam splitters. We will firstly analyse these advantages from the perspective of what is optimal for the two-photon scattershot source, which consists of an array of non-linear squeezing crystals and photon detectors. Each crystal, with every pump pulse, has a chance of generating a pair of photons through spontaneous parametric down conversion (SPDC). One of the photons is funnelled to a photon detector, which heralds the existence of the other photon, as shown in Fig.~\ref{fig:f4}\textbf{a}. This heralded photon can then be run through an optical circuit $U$, before also being measured by detectors. This useful set up allows us to record how many, and in which channels, photons enter into $U$. Note that it is possible for multiple crystals to give off photons at the same time, thus allowing $U$ to perform multi-photon dynamics between initially dispersed photons. We only accept situations where two photons are heralded at the same time in two separate modes, as shown in Fig.~\ref{fig:f4}\textbf{b}. 

We define the probability of success $\mathrm{P}_n(U)$ as the probability that the scattershot source generates the correct number and configuration of photons for input into a given system $U$. So that the comparisons will be fair, we will only consider systems which can accept input from a fixed $n$ number of these non-linear crystals. With all other things being equal, a higher probability of success $\mathrm{P}_n(U')<\mathrm{P}_n(U)$ essentially means that model $U$ has a faster sampling rate compared to model $U'$. Note that an increase in $n$ also would correspond to an increase sampling rate, though the scaling of this increase will differ depending on the particular model $U$ chosen. These points will be properly quantified in the following section. 
 
Each non-linear crystal through SPDC produces the following pairs of photons in a squeezed state  
\begin{align} 
    |\psi\rangle_q = \sqrt{1-\chi^2}\sum_{p=0}^\infty \chi^p|pp\rangle_q, 
\end{align} 
where $0\leq\chi\leq1$ is the parameter that determines the strength of the squeezing~\cite{lund2014boson}. Hence the total state for an array of $n$ non-linear squeezers is the following 
\begin{align}
    |\Psi\rangle=\bigotimes^n_{q=1}|\psi\rangle_q=\left(1-\chi^2\right)^{\frac{n}{2}}\prod_{q=1}^n\sum_{p=0}^\infty \chi^p|pp\rangle_q. 
\end{align}
Now, the probability of heralding two photons and nothing elsewhere, say in the first two modes $|\Phi\rangle=|11\rangle_1\bigotimes|11\rangle_2\bigotimes^n_{p=3}|00\rangle_p$, is given by 
\begin{align}
    \left|\langle\Phi|\Psi\rangle \right|^2 = (1-\chi^2)^n\chi^4.  \label{eq:2_p2ph} 
\end{align}
However, in our $D_m$, we are allowed to herald two photons anywhere in the $m=n$ possible modes, in which there are $\binom{m}{2} = m(m-1)/2$ possible allowed inputs. Hence the net probability of success is given by 
\begin{align}
    \mathrm{P}_n(D_n)=(1-\chi^2)^n\chi^4\frac{n(n-1)}{2}. 
\end{align}
On the other hand, if our system is a simple linear array of $n/2$ beam splitters $L_n$, the two photons must be heralded in the correct ports, as shown in Fig.~\ref{fig:f4}\textbf{c}. There are only $n/2$ acceptable inputs hence the total probability of success is 
\begin{align}
    \mathrm{P}_n(L_n)=(1-\chi^2)^n\chi^4\frac{n}{2}. 
\end{align}
It is clear that for all possible sizes of the system that $D_n$ always provides a better probability of success compared to $L_n$, and this ratio increases with increasing $n$. Hence the designed MHD $D_n$ provides a much faster sampling rate compared to $L_n$. 

We also consider modifying the source itself, as shown in Fig.~\ref{fig:f4}\textbf{d}. In this situation, we don't funnel one half of the photons into heralding detectors, instead the two output ports of the non-linear crystal are connected directly to the two input ports of a beam splitter. In that case, we only need a single crystal to generate a $|11\rangle$ pair, which simply changes the non-linear factor in Eq.~\eqref{eq:2_p2ph} from $\chi^4$ to $\chi^2$. Since there will be a total of $n$ squeezing crystals and $n$ beam splitters in the system $L_{2n}'$, this means the net probability of success is 
\begin{align}
    \mathrm{P}_n(L_{2n}') = (1-\chi^2)^n\chi^2n. 
\end{align}
It is apparent, for smaller $n$ values and system sizes, this $L_{2n}'$ configuration has a higher sampling rate over $D_n$. However, above $n = 2/\chi^2 + 1$ amount of squeezers, $D_n$ once again provides a better sampling rate over $L_{2n}'$, where the ratio of this advantage grows as $n$ increases. Note that this modified source has the significant limitation in that the input is not heralded; we don't know that a squeezer crystal has emitted paired photons until after it has been detected after passing through the $L_{2n}'$ system. In contrast, the scattershot source allows us to know about the input beforehand. Therefore, we have knowledge of the output state and can in principle hold it in quantum memory as a known resource state, before choosing to do either further quantum processing or detection.

We note that $D_n$ has another advantage over both $L'_{2n}$ and $L_n$ in that it has the ability to count the number of photons without photon number resolving detectors. If we are restricted to on/off bucket photon detectors, we will still be able to measure say two photons landing in $A$ because they can land separately in any detector in group $A$. It is clear that $L'_{2n}$ and $L_n$ can't do this because they are composed of separate beam splitter systems stacked on top of each other. Furthermore, due to the mode mixing in the $D_n$ system, it can distinguish the instances where the source generates the incorrect input of more than two photons. This counting and error detection attributes means that the MHD can provide more accurate results compared to other systems. 

The above computational gains come at a cost in terms of the physical resources and amount of optical components needed to make this device. All $D_n$ can be decomposed into at most $n(n-1)/2$ two-level unitary matrices or beam splitters~\cite{reck1994experimental}. As an example, we show the resulting circuit for $D_4$ in Fig.~\ref{fig:f4}\textbf{e}, determined using the particular decomposition method given in~\cite{nielsen2010quantum}. Note that we factored out extra $\pi$ phase shifts so that the two-level beam splitters given in this diagram are all in the form 
\begin{align}
    \begin{pmatrix} 
        \sqrt{1-\eta_i} & \sqrt{\eta_i} \\ 
        \sqrt{\eta_i} & -\sqrt{1-\eta_i} 
    \end{pmatrix},
\end{align}
where the transmission ratios are given by $\eta_1=\sin^2\theta/[2+\cos(2\theta)]$, $\eta_2=2\sin^2\theta/[5+\cos(2\theta)]$, and $\eta_3=\sin^2\theta/3$. 

\section{Conclusion} \label{sec:MHD_con}

We have shown that there exists a multimode HOM device $D_m(\theta)$ which can replicate the two-photon statistics of a beam splitter, irrespective of where these two photons enter amongst the system's $m$ modes. The fact that this circuit can display exactly the same quantum mechanical number statistics, invariant on the input modes, makes it an interesting case study. We firstly claim that this circuit can be generated from a skew-symmetric orthogonal matrix, whose off-diagonal elements are the same magnitude. We then show from these properties that at particular critical transmission ratios $\theta$, it can be tuned to replicate either the HOM dip or a 100\% coincidence rate. Furthermore, we show that the number statistics will decrease continuously with increasing $\theta$ between these two critical points, thus allowing the possibility to map it to the statistics profile of a typical beam splitter. Finally, we show that that this device provides experimental advantages in terms of higher sampling rate with better accuracy compared to other more straightforward systems. 

\begin{acknowledgments}
This research was supported by the Australian Research Council Centre
of Excellence for Quantum Computation and Communication
Technology (Project No. CE110001027).
\end{acknowledgments}

\bibliography{paper}

\appendix*

\section{Direct Calculation of the Total Two Photon Bunching Probability} \label{sec:MHD_app2bunching}

We can directly calculate the total probability of the two photons landing in the $A$ labelled detectors, by considering the squared permanent of relevant rows in Eq.~\eqref{eq:MHD_cicj}. There are $\binom{m/2-1}{2} = \left(m/2-1\right)\left(m/2-2\right)/2$ possible outcomes in which two photons land in different detectors $r_a\neq r_{a'}$, but the same group $r_a,r_{a'}\in A$, with a probability of 
\begin{align}
    \mathbb{P}_m(r_a r_{a'} |c_i c_j;\theta) &= \left( s_{a,i}'s_{a',j}'\frac{\sin^2 \theta}{m-1} + s_{a',i}'s_{a,j}'\frac{\sin^2 \theta}{m-1} \right)^2, \nonumber \\ 
    &= \frac{4\sin^4\theta}{(m-1)^2},\hspace{5 mm} \forall a\neq i,a'\neq i. \nonumber
\end{align}
It is the case that $s_{a,i}'s_{a',j}'=s_{a',i}'s_{a,j}'$ since the two output detectors belong to the same grouping. There also exists $m/2-1$ coincidence probabilities of the form 
\begin{align}
    \mathbb{P}_m(r_a r_i |c_i c_j;\theta) &= \left(s_{a,j}'\frac{\cos \theta \sin \theta}{\sqrt{m-1}} + s_{i,j}'s_{a,i}'\frac{\sin^2 \theta}{m-1} \right)^2, \nonumber \\ 
    &= \left(\frac{\cos \theta \sin \theta}{\sqrt{m-1}} + \frac{\sin^2 \theta}{m-1} \right)^2,\hspace{5 mm} \forall a\neq i. \nonumber
\end{align}
We also need to consider the cases where the two photons land in the same detector $\mathbb{P}_m(r_a^2|c_i c_j;\theta)$. We note that in these cases, we need to divide the permanent squared by 2, due to considerations of bunching~\cite{gard2015introduction,lund2017quantum}. There are $m/2-1$ of the form 
\begin{align}
    \mathbb{P}_m(r_a^2 |c_i c_j;\theta) &= \frac{1}{2}\left( s_{a,i}'s_{a,j}'\frac{\sin^2 \theta}{m-1} + s_{a,i}'s_{a,j}'\frac{\sin^2 \theta}{m-1} \right)^2, \nonumber \\ 
    &= \frac{2\sin^4\theta}{(m-1)^2},\hspace{5 mm} \forall a\neq i. \nonumber
\end{align}
Finally, there is 1 individual probability of the form 
\begin{align}
    \mathbb{P}_m(r_i^2 |c_i c_j;\theta) &= \frac{1}{2}\left( s_{i,j}'\frac{\cos\theta \sin\theta}{\sqrt{m-1}} + s_{i,j}'\frac{\cos\theta \sin\theta}{\sqrt{m-1}} \right)^2, \nonumber \\ 
    &= \frac{2\cos^2\theta\sin^2\theta}{m-1}.\nonumber
\end{align}
Now, we can add up all these individual probabilities to get the total probability of two photons landing group $A$ detectors
\begin{align}
    \mathbb{P}_m(A^2 |c_i c_j;\theta) &= \frac{1}{2}\left(\frac{m}{2}-1\right)\left(\frac{m}{2}-2\right)\frac{4\sin^4\theta}{(m-1)^2} \nonumber \\ 
    &\hspace{5 mm}+ \left(\frac{m}{2}-1\right)\left(\frac{\cos \theta \sin \theta}{\sqrt{m-1}} + \frac{\sin^2 \theta}{m-1} \right)^2 \nonumber \\
    &\hspace{5 mm}+ \left(\frac{m}{2}-1\right)\frac{2\sin^4\theta}{(m-1)^2}+\frac{2\cos^2\theta\sin^2\theta}{m-1}, \nonumber \\ 
    &= \frac{m-2}{2(m-1)}\sin^4 \theta + \frac{m-2}{(m-1)^{3/2}}\cos\theta\sin^3\theta \nonumber \\ 
    &\hspace{5 mm}+ \frac{m+2}{2(m-1)}\cos^2\theta\sin^2\theta. \nonumber
\end{align}
This probability expression is consistent with Eq.~\eqref{eq:MHD_pA2} and the total coincidence probability.  

\end{document}